\documentclass[pra,superscriptaddress,twocolumn,showpacs,preprintnumbers,amsmath,amssymb]{revtex4-1}

\usepackage{graphicx}
\usepackage{dcolumn}
\usepackage{bm}

\begin{document}

\title{Magnetic phases of mass- and population-imbalanced ultracold fermionic mixtures in optical lattices}

\author{Andrii Sotnikov}
\affiliation{Institut f\"ur Theoretische Physik, Goethe-Universit\"at, 60438 Frankfurt/Main, Germany}
\author{Michiel Snoek}
\affiliation{Institute for Theoretical Physics, Universiteit van Amsterdam, 1090 GL Amsterdam, The Netherlands}
\author{Walter Hofstetter}
\affiliation{Institut f\"ur Theoretische Physik, Goethe-Universit\"at, 60438 Frankfurt/Main, Germany}

\date{\today}

\begin{abstract}
We study magnetic phases of two-component mixtures of ultracold fermions with repulsive interactions in optical lattices in the presence of both hopping and population imbalance by means of dynamical mean-field theory (DMFT). It is shown that these mixtures can have easy-axis antiferromagnetic, ferrimagnetic, charge-density wave, and canted-antiferromagnetic order or be unordered depending on parameters of the system. We study the resulting phase diagram in detail and investigate the stability of the different phases with respect to thermal fluctuations. We also perform a quantitative analysis for a gas confined in a harmonic trap, both within the local density approximation and using a full real-space generalization of DMFT. 
\end{abstract}

\maketitle

\section{Introduction}
At the present moment, ultracold atoms in optical lattices can be considered as one of the most powerful tools for testing Hubbard-type models \cite{Blo2008RMP}. In these systems parameters like the hopping amplitude, interaction strength, number and type of atomic species, lattice geometry, and dimensionality can be tuned in a wide range, which makes it experimentally possible to verify predictions of theoretical models in different regimes. Despite incredible progress in the field of ultracold atoms, and, in particular, observation of short-range magnetic correlations in a recent experiment \cite{Greif2012pre}, probing quantum magnetic phenomena still remains a challenging goal for experimentalists in this field, because the range of entropies (and temperatures) that are currently accessible is still several times higher than the upper limit for observing magnetic long-range order \cite{Jor2010PRL}. Once this important problem has been solved, it is believed that ultracold atomic systems can give 
significant insight into high-temperature superconductivity \cite{Hof2002PRL} and will be highly promising for quantum simulations \cite{Simon2011Nat} in general.

In this paper, we focus on a generalization of the Fermi-Hubbard model by considering two-component mixtures with repulsive interactions, which have an imbalance both in the hopping amplitude (corresponding to different effective masses) and in the population of the two components. In ultracold atomic mixtures, this generalization was theoretically studied for the case of attractive interactions in the context of competing superfluid and density-wave ground states \cite{Dao2007PRB}, Fulde-Ferrell-Larkin-Ovchinnikov superfluidity and other long-range ordered phases \cite{Dalmonte2012PRA,Wang2013PRA}. For the case of repulsive interactions, it was investigated in the context of itinerant ferromagnetism (Stoner instability) \cite{Keyserlingk2011PRA, Cui2013PRL}. Antiferromagnetic phases in ultracold imbalanced Fermi-Fermi mixtures with moderately strong interactions were studied only separately in the case of population \cite{Gottwald2009PRA,Wunsch2010PRA,Koetsier2010PRA,Snoek2011PRB} or mass imbalance 
\cite{Caz2005PRL,Sotnikov2012PRL,Winograd2012PRB}. 
Evidently, so far the magnetic ordering phenomena in the case, where both types of imbalance are present, have not been considered. Our paper aims at bridging this gap.

The paper is organized as follows. In Sec.~\ref{sec.2} we introduce the model and derive an effective Hamiltonian, which allows us to interpret our numerical results. In Sec.~\ref{sec.3} we briefly outline the main steps in the generalization of our numerical approach, dynamical mean-field theory (DMFT) \cite{Geo1996RMP}, that are important for a proper account of magnetic ordering effects in the case of mass and population imbalance. Section~\ref{sec.4} is devoted to our results and their discussion.

\section{Model}\label{sec.2}
We consider a Fermi-Hubbard Hamiltonian of the following type:
\begin{eqnarray}
\mathcal{\hat{H}}=&&
-\sum\limits_{\langle ij\rangle}\sum\limits_{\sigma} t_\sigma (\hat{c}^\dag_{i\sigma}\hat{c}_{j\sigma}+{\rm h.c.})
+U\sum\limits_{i}\hat{n}_{i\uparrow}\hat{n}_{i\downarrow}
\nonumber\\
&&+\sum\limits_{i}\sum\limits_{\sigma}(V_i-\mu_\sigma)\hat{n}_{i\sigma},
\label{eq.1}
\end{eqnarray}
where $t_\sigma$ is the hopping amplitude of fermionic species $\sigma=\{\uparrow,\downarrow\}$,
$\hat{c}^\dag_{i\sigma}$ ($\hat{c}_{i\sigma}$) is the corresponding creation (annihilation) operator of species $\sigma$ at the lattice site~$i$,
the notation $\langle ij\rangle$ indicates a summation over nearest-neighbor sites, and $U$ is the magnitude of the on-site repulsive ($U>0$) interaction of the two different species with corresponding densities $\hat{n}_{i\uparrow}$ and $\hat{n}_{i\downarrow}$  ($\hat{n}_{i\sigma}=\hat{c}^\dag_{i\sigma}\hat{c}_{i\sigma}$). 
In the last term, $V_i$ is the external (e.g., harmonic) potential at lattice site $i$, and $\mu_\sigma$ is the chemical potential of species $\sigma$. 
Note that we have taken the harmonic potential to be independent of the atomic species.
The Hamiltonian~(\ref{eq.1}) implies a single-band approximation; in other words, we consider the case of a sufficiently strong lattice potential, $V_{\text{lat}}\gtrsim 5E_r$. 

It should be noted that within this model, the so-called mass imbalance depends only on the hopping amplitudes $t_\sigma$, which for sufficiently deep lattices are given by \cite{Zwe2003JOB}
\begin{equation}\label{eq.2}
 t_\sigma \approx \frac{4}{\sqrt\pi}E_{r\sigma} v_\sigma^{3/4}
 \exp\bigl(-2\sqrt{v_\sigma}\bigr),
\end{equation}
where $v_\sigma =V_{\text{lat}}^{(\sigma)}/E_{r\sigma}$, $E_{r\sigma}={\hbar^2k^2}/{2m_\sigma}$ is the recoil energy, $k$ is the wave number determined by the wavelength of the laser forming the optical lattice, and $m_\sigma$ is the mass of species $\sigma$.
The amplitude $V_{\text{lat}}^{(\sigma)}$ of the lattice potential can be different for the two components $V_{\text{lat}}^{(\uparrow)}\neq V_{\text{lat}}^{(\downarrow)}$, which results in the possibility to realize an imbalance in the hopping amplitude even for different hyperfine states of the same atom ($m_\uparrow=m_\downarrow$) \cite{Man2003PRL}.
As for the population imbalance, its magnitude depends not only on the chemical potentials~$\mu_\sigma$, but also on other parameters, including the mass imbalance $\Delta t=(t_\uparrow-t_\downarrow) /(t_\uparrow+t_\downarrow)$. In order to characterize the population imbalance quantitatively, we introduce the polarization $P=(N_\uparrow-N_\downarrow)/(N_\uparrow+N_\downarrow)$, where $N_\sigma$ is the total number of particles of type $\sigma$ in the system.

It is important to mention that the introduced mass-imbalance parameter $\Delta t$ can be experimentally tuned in a wide range. For example, according to Ref.~\cite{Dalmonte2012PRA}, for a $^6$Li-$^{40}$K mixture it can be effectively tuned from 0.3 to 0.85 by varying the intensity (in the range of 1~W) and detuning (in the range of 2~nm from the magic wavelength) of the laser beams forming the optical lattice. 
Other systems where hopping imbalance can be realized and tuned in different ranges include the $^{171}$Yb-$^{173}$Yb mixture \cite{Tai2010PRL}, spin-dependent optical lattices for homonuclear mixtures and mixtures of alkali-metal and alkaline-earth fermionic atoms. By approaching the limit of large imbalance, $\Delta t\rightarrow1$, these systems even allow for studies of the Falicov-Kimball model \cite{Freericks2003RMP} that is used to model certain solid-state materials. 

By using  the Schrieffer-Wolff transformation in the limit $t_{\sigma}\ll U$ near half-filling, $n_{i\uparrow}+n_{i\downarrow}\approx1$, we can map the Hamiltonian~(\ref{eq.1}) to an effective spin model \cite{Kuk2003PRL,Alt2003NJP}.
For the system under study, this transformation results in the anisotropic Heisenberg (XXZ) model,
\begin{eqnarray}
 \mathcal{\hat{H}}_{\textrm{eff}}&=&J_{\parallel}\sum_{\langle ij\rangle}\hat{S}_{i}^{Z}\hat{S}_{j}^{Z}
 +J_{\perp}\sum_{\langle ij\rangle}(\hat{S}_{i}^{X}\hat{S}_{j}^{X}+\hat{S}_{i}^{Y}\hat{S}_{j}^{Y})
 \nonumber
\\
 && - \Delta\mu\sum_i \hat{S}_{i}^{Z},
 \label{eq.3}
\end{eqnarray}
with coupling constants 
\begin{equation}\label{eq.4}
 J_{\parallel}= 2(t_\uparrow^2+t_\downarrow^2)/U,\quad
 J_{\perp}=4t_\uparrow t_\downarrow/U,
\end{equation}
and a chemical potential difference $\Delta\mu = \mu_\uparrow-\mu_\downarrow$ that plays the role of an external magnetic field.
$\hat{S}^R_{i}$ ($R=\{X,Y,Z\}$) are usual spin-1/2 operators on the lattice site $i$, $\hat{S}^R_{i}= \frac{1}{2}\hat{c}^\dag_{i\alpha}\sigma^R_{\alpha\beta}\hat{c}_{i\beta}$, where $\sigma^R$ are the Pauli matrices. It should be noted that here and below, all ``magnetic'' characteristics refer to a pseudospin made of two different species, thus the anisotropy in the spin model is not related to a spatial direction or quantization axis in the original lattice.

The anisotropy in the Hamiltonian~(\ref{eq.3}) originates from the imbalance in hopping amplitudes. This reduces the initial $SU(2)$ rotational spin symmetry of the balanced mixture to the lower $\mathbb{Z}_2\times U(1)$ spin symmetry, where the symmetry group $\mathbb{Z}_2$ corresponds to reflections about the $XY$ plane [Ising type; first term in Eq.~(\ref{eq.3})] and $U(1)$ corresponds to spin rotations in the $XY$ plane [second term in Eq.~(\ref{eq.3})]. The symmetry of the model is further reduced to $U(1)$ if one has a nonzero chemical potential difference $\Delta\mu$. 

According to Eqs.~(\ref{eq.4}), we note that the coupling $J_{\parallel}$ is always larger than or equal to $J_{\perp}$. Hence, from Eq.~(\ref{eq.3}) one concludes that at $\Delta\mu=0$ and $\Delta t\neq0$ the ground state of the system is an easy-axis antiferromagnet (Ising-type; Z-AF). As was pointed out in Ref.~\cite{Sotnikov2012PRL}, the excitation spectrum is gapped in this case and the Z-AF phase is also permitted in low dimensions ($d<3$; as the Mermin-Wagner theorem applies only to continuous symmetries). Evidently, this also holds for a nonzero but small $\Delta\mu < (J_{\parallel}-J_{\perp})$, when the system's ground state corresponds to a ferrimagnet (Z-AF with an additional net magnetization in the $Z$ direction). In the opposite case, $\Delta\mu\neq0$ and $\Delta t=0$, the ground state is a canted antiferromagnet (easy-plane antiferromagnet with a net magnetization in $Z$ direction), which obeys the Mermin-Wagner theorem and was studied in Refs.~\cite{Gottwald2009PRA,Wunsch2010PRA,Koetsier2010PRA,
Snoek2011PRB}. Therefore, in the region of intermediate imbalances, $\Delta\mu\sim(J_{\parallel}-J_{\perp})$, one should expect a phase transition between these two different types of magnetic ordering.

The effective Hamiltonian~(\ref{eq.3}) gives a good understanding of the types of ordered phases arising in the system under study. 
However, in order to have a more complete picture of the structure and quantitative characteristics of the magnetically ordered phases arising in optical lattices at nonzero temperature and governed by the Hamiltonian~(\ref{eq.1}), it is necessary to use nonperturbative numerical approaches. In this article, we apply DMFT, which is well suited for the description of long-range ordered phases in high-dimensional systems and is able to fully capture effects of inhomogeneity and finite-size that are usually present in optical lattices.

\section{Method}\label{sec.3}
Dynamical mean-field theory \cite{Geo1996RMP} is an approach that makes it possible to bridge two limits on the lattice: the nearly free fermion gas and the strongly interacting (atomic) limit. This method maps the lattice problem (which is, in general, intractable) to an impurity problem (which is chosen to be numerically solvable), thus substituting the full action with an effective one. Despite the fact that it is a nonperturbative approach, it is still an approximate method, since it treats the lattice self-energy as a local (which for homogeneous systems means: momentum-independent) quantity, thus neglecting nonlocal quantum fluctuations. DMFT becomes exact in the limit of infinite dimensions, $d=\infty$ (i.e., large coordination number $z\gg1$). Although it is not an exact method in the case of square and cubic lattice geometries ($z=4$ and $z=6$, respectively), results obtained by DMFT can be used as a reference point both for experiments and for more sophisticated methods, such as quantum Monte Carlo simulations, 
which could be 
computationally rather demanding due to the presence of a sign problem in the case under study (for recent results, possibilities, and limitations, see \cite{Kozik2012pre} and references therein).

\subsection{Impurity model and solver}
Most commonly, for solving the auxiliary impurity problem in DMFT, the lattice model~(\ref{eq.1}) is mapped onto an Anderson impurity model (AIM). 
The model Hamiltonian contains the full local physics of the lattice problem, but the nonlocal terms are represented by noninteracting bath degrees of freedom.
In the case under study this Hamiltonian can be written in the following form:
\begin{eqnarray}
\hat{\cal H}_{\text{AIM}}
= & &
  \sum_{l=2}^{n_s}\sum_{\sigma} \varepsilon_{l\sigma} \hat{a}_{l\sigma}^\dag \hat{a}_{l\sigma} 
+ \sum_{l=2}^{n_s}\sum_{\sigma} V_{l\sigma} (\hat{a}_{l\sigma}^\dag \hat{d}_{\sigma} + {\rm h.c.})
\nonumber
\\
&+&
  \sum_{l=2}^{n_s}\sum_{\sigma} \Delta_l \hat{a}_{l\sigma}^\dagger \hat{a}_{l\bar{\sigma}}
+ \sum_{l=2}^{n_s}\sum_{\sigma} W_{l\sigma} (\hat{a}_{l\sigma}^\dagger \hat{d}_{\bar{\sigma}} + {\rm h.c.})
\nonumber
\\
&+& U\hat{n}_{d\uparrow}\hat{n}_{d\downarrow}
 - \sum_{\sigma}\mu_{\sigma}\hat{n}_{d\sigma}
 - \mu_{\uparrow\downarrow}^{(i)}\sum_{\sigma}\hat{d}_{\sigma}^\dag \hat{d}_{\bar{\sigma}},
\label{eq.5}
\end{eqnarray}
where $\sigma$ and its opposite $\bar{\sigma}$ represent the spin indices, $\sigma=\{\uparrow,\downarrow\}$, and the index $l=\{2,\ldots,n_s\}$ labels the number of the bath's orbital in AIM with $n_s$ being the cut-off number. 
In our calculations we use the exact diagonalization (ED) solver \cite{Caffarel1994PRL} with $n_s=5$, such that there are four orbitals.
The operators $\hat{a}^\dag_{l\sigma}$ ($\hat{a}_{l\sigma}$) and $\hat{d}^\dag_{\sigma}$ ($\hat{d}_{\sigma}$) are the creation (annihilation) operators of electrons on the bath's orbital $l$ and the impurity, respectively; the quantities $\varepsilon_{l\sigma}$, $V_{l\sigma}$, $\Delta_l$ and $W_{l\sigma}$ are the so-called Anderson parameters that set the amplitude of different processes in this model.
In particular, the two terms in the first line of Eq.~(\ref{eq.5}) correspond to the energies of electrons in the bath and the hybridization between the bath and the impurity, respectively. 
The second line represents the anomalous terms that are important for obtaining quantities corresponding to magnetic ordering in the $XY$ plane. The terms in the last line have a direct correspondence to the Hamiltonian~(\ref{eq.1}), except the last (auxiliary) term, which is used in calculations as a small initial perturbation to break the remaining $U(1)$ symmetry of the spin model~(\ref{eq.3}).

Using the standard technique (see, e.g., Ref.~\cite{Stoof2009}) of eliminating bath degrees of freedom in the effective action corresponding to the Hamiltonian~(\ref{eq.5}), we find analytical expressions for the Weiss Green's functions, which represent effective dynamical fields acting on the impurity site:
\begin{eqnarray}
{\cal G}_\sigma^{-1}(i\omega_n) 
=
i\omega_n + \mu_\sigma 
- \sum_{l=2}^{n_s}
K_l^{-1}\left[
V_{l\sigma}^2(i\omega_n-\varepsilon_{l\bar{\sigma}})
\right. \nonumber
\\
\left.
+ 2V_{l\sigma}W_{l\sigma}\Delta_l + W_{l\sigma}^2(i\omega_n-\varepsilon_{l\sigma})
\right], \nonumber
\\
{\cal G}_{\uparrow\downarrow}^{-1}(i\omega_n) 
=
\mu^{(i)}_{\uparrow\downarrow}- \sum_{l=2}^{n_s}
K_l^{-1}\left[
V_{l\uparrow}W_{l\downarrow}(i\omega_n-\varepsilon_{l\downarrow})
\right. \nonumber
\\
\left. 
+ (V_{l\uparrow}V_{l\downarrow}+W_{l\uparrow}W_{l\downarrow})\Delta_l 
+ V_{l\downarrow}W_{l\uparrow}(i\omega_n-\varepsilon_{l\uparrow}) 
\right], 
\label{eq.6}
\end{eqnarray}
where $K_l=(i\omega_n-\varepsilon_{l\uparrow})(i\omega_n-\varepsilon_{l\downarrow})-\Delta_l^2$,
$\omega_n = \pi(2n+1)/\beta$ is the Matsubara frequency and $\beta$ is the inverse temperature, $\beta=1/T$ (we use units such that $k_B=1$).

Within the ED solver, the basis states of the finite-dimensional Hilbert space are given by
\begin{equation}
 |n_1^\uparrow,n_2^\uparrow,\ldots,n_{n_s}^\uparrow\rangle
 |n_1^\downarrow,n_2^\downarrow,\ldots,n_{n_s}^\downarrow\rangle
\end{equation}
with $n_p^\sigma=0,1$ and $\sum_p n_p^\sigma \equiv n^\sigma$.
Note that the anomalous terms in Eq.~(\ref{eq.5}) mix the sectors $n^\uparrow$ and $n^\downarrow$ (i.e., the magnetization $s_z$ is not conserved), which therefore cannot be diagonalized independently.
Although the total charge $n=n^\uparrow+n^\downarrow$ is still conserved, this leads to a significant increase of the numerical effort in diagonalization and subsequent calculations of the corresponding Green's functions. At finite temperature, these are calculated from the full set of eigenstates~$|i\rangle$ (with eigenvalues $E_i$) according to
\begin{eqnarray}
 G_{\sigma_1\sigma_2}(i\omega_n) =&& \frac{1}{\cal Z}\sum_{i,j}
 \frac{\langle i|\hat{d}_{\sigma_1}|j\rangle\langle j|\hat{d}^\dag_{\sigma_2}|i\rangle}{E_i-E_j-i\omega_n}
 \nonumber
\\  
 &&\times\left( e^{-\beta E_i} + e^{-\beta E_j} \right),
\label{eq.8}
\\
 F_{\sigma_1\sigma_2}(i\omega_n) =&& \frac{1}{\cal Z}\sum_{i,j}
 \frac{\langle i|\hat{d}_{\sigma_1}\hat{d}^\dag_{\bar{\sigma}_1}\hat{d}_{\bar{\sigma}_1}|j\rangle\langle j|\hat{d}^\dag_{\sigma_2}|i\rangle}{E_i-E_j-i\omega_n}
 \nonumber
\\  
 &&\times\left( e^{-\beta E_i} + e^{-\beta E_j} \right),
\label{eq.9}
\end{eqnarray}
where ${\cal Z}=\sum_i e^{-\beta E_i}$ is the partition function. Next, following Ref.~\cite{Bulla1998JP}, the self-energies can be defined as 
\begin{eqnarray}
 \Sigma_{\sigma\sigma} &=& U \frac{F_{\sigma\sigma} G_{\bar{\sigma}\bar{\sigma}} - F_{\sigma\bar{\sigma}} G_{\sigma\bar{\sigma}}}
 {G_{\sigma\sigma} G_{\bar{\sigma}\bar{\sigma}} - G_{\sigma\bar{\sigma}}^2},
\label{eq.10}
\\
 \Sigma_{\sigma\bar{\sigma}} &=& U \frac{F_{\sigma\bar{\sigma}} G_{\sigma\sigma} - F_{\sigma\sigma} G_{\sigma\bar{\sigma}} }
 {G_{\sigma\sigma} G_{\bar{\sigma}\bar{\sigma}} - G_{\sigma\bar{\sigma}}^2}.
\label{eq.11}
\end{eqnarray}

In practice, we solve the impurity problem by obtaining the quantities (\ref{eq.8})--(\ref{eq.11}) for given parameters $U$, $\mu_\sigma$, and $\beta$ from the original lattice problem (\ref{eq.1}) and for a particular set of auxiliary Anderson parameters $\{\varepsilon_{l\sigma}, V_{l\sigma}, W_{l\sigma}, \Delta_l \}$, which is updated in each DMFT iteration.
The self-energies (\ref{eq.10}) and (\ref{eq.11}) then allow us to calculate the Green's functions corresponding to the initial lattice problem. Below we consider two main approaches for evaluating these Green's functions: (i) two-sublattice DMFT, which is important for obtaining phase diagrams for homogeneous (infinite) systems with magnetic order and can also be used in combination with a local-density approximation (LDA) to analyze trapped gases, and (ii) real-space DMFT, which describes finite inhomogeneous (trapped) systems without further approximations.

\subsection{Two-sublattice DMFT}\label{sec.3.2}
Bipartite structures, such as states with antiferromagnetic order, can be described in an appropriate way by introducing two sublattices. 
Within the DMFT approach one then needs to solve the impurity problem twice on two adjacent sites of the original lattice.
It is important to note that here, in contrast to the case of balanced mixtures, the observables corresponding to different sublattices (denoted below by $s=1,2$) cannot be directly associated with observables corresponding to different species (denoted by $\sigma=\uparrow,\downarrow$). 
Hence, we define the lattice Green's functions in the following (generalized) way:
\begin{equation} \label{eq.12}
 G^{(s)}_{\sigma_1\sigma_2}(i\omega_n) = 
 \int_{-z}^{z} d\epsilon\, {D(\epsilon)}[{\bf A}^{-1}(\epsilon)]^{(s)}_{\sigma_1\sigma_2}
\end{equation}
with
\begin{equation}
 {\bf A}=\left(
 \begin{array}{cccc}
  \zeta_{\uparrow\uparrow}^{(1)} & \zeta_{\downarrow\uparrow}^{(1)} & -t_\uparrow\epsilon & 0\\
  \zeta_{\uparrow\downarrow}^{(1)} & \zeta_{\downarrow\downarrow}^{(1)}   & 0 & -t_\downarrow\epsilon\\
  -t_\uparrow\epsilon & 0 & \zeta_{\uparrow\uparrow}^{(2)} & \zeta_{\downarrow\uparrow}^{(2)}\\
  0 & -t_\downarrow\epsilon & \zeta_{\uparrow\downarrow}^{(2)} & \zeta_{\downarrow\downarrow}^{(2)}  \\
 \end{array}
\right)
\end{equation}
and
\begin{eqnarray}
 &&\zeta_{\sigma\sigma}^{(s)}=i\omega_n+\mu_\sigma - \Sigma_{\sigma\sigma}^{(s)}(i\omega_n),
 \label{eq.14}
 \\
 &&\zeta_{\sigma\bar{\sigma}}^{(s)}=\mu_{\sigma\bar{\sigma}}^{(s)} - \Sigma_{\sigma\bar{\sigma}}^{(s)}(i\omega_n),
 \label{eq.15}
\end{eqnarray}
where $z$ denotes the lattice coordination number ($z=4$ and $z=6$ for square and cubic lattices, respectively), $D(\epsilon)$ is the normalized density of states, $\int_{-z}^z d\epsilon\, {D(\epsilon)}=1$, the explicit form of which is known for a particular lattice geometry. 
The local self-energies appearing in Eqs.~(\ref{eq.14}) and (\ref{eq.15}) are taken from the impurity solver [see Eqs.~(\ref{eq.10}) and (\ref{eq.11})].

To complete the self-consistency equations of the DMFT scheme, we define the Weiss Green's functions from the Dyson equation,
\begin{equation}
  [{\cal G}^{(s)}(i\omega_n)]^{-1}_{\sigma_1\sigma_2} 
= [{\bf G}^{(s)}(i\omega_n)]^{-1}_{\sigma_1\sigma_2} + \Sigma_{\sigma_1\sigma_2}^{(s)}(i\omega_n).
\end{equation}
where ${\bf G}^{(s)}$ is a $2\times2$ block of the matrix~(\ref{eq.12}), the inverse of which is taken separately for different sublattices $s=1,2$.
By using the obtained Weiss Green's functions in the minimization procedure [according to Eqs.~(\ref{eq.6}) and applying a conjugate gradient method], we find a new set of Anderson parameters, which is then used in a subsequent DMFT iteration. These iterations are performed until final convergence, i.e., until the initial and final Weiss Green's functions coincide.

When an inhomogeneous system is considered, the two-sublattice DMFT introduced above can be used in combination with LDA.
The main advantage of LDA+DMFT is that this approach allows to consider large systems in three dimensions, as the numerical effort scales approximately linearly with the system size for axial-symmetric trapping potentials. The drawback of this approach is that it fails to reproduce the detailed structure close to the boundaries of the ordered phases; i.e., it does not account for a possible proximity effect. The mentioned effect can be accounted for within another generalization: the real-space DMFT, which was first introduced in Refs.~\cite{Hel2008PRL,Sno2008NJP}.

\subsection{Real-space DMFT}
The main idea of real-space DMFT (R-DMFT) is not to divide the lattice problem into several sublattices, but to solve the impurity problem on each lattice site corresponding to the original finite-size system directly. Then, after we obtain the self-energies $\Sigma^{(i)}_{\sigma_1\sigma_2}(i\omega_n)$ [see Eqs.~(\ref{eq.10}) and (\ref{eq.11})] for each lattice site~$i=\{1,...,N\}$, we collect them in the real-space matrix consisting of inverse local Green's functions and hopping elements,
\begin{equation}\label{eq.17}
 {\bf G}^{-1}=\left(
\begin{array}{ccccccc}
\zeta^{(1)}_{\uparrow\uparrow} &  &  &  &  &  &\\
\zeta^{(1)}_{\uparrow\downarrow} & \zeta^{(1)}_{\downarrow\downarrow}  &  &  &  &  &\\
t_{\uparrow} & 0 & \zeta^{(2)}_{\uparrow\uparrow} &  &  &  &\\
0 & t_{\downarrow} & \zeta^{(2)}_{\uparrow\downarrow} & \zeta^{(2)}_{\downarrow\downarrow}  &  &  &\\
0 & 0 & t_{\uparrow} & 0 & \zeta^{(3)}_{\uparrow\uparrow} &  &\\
0 & 0 & 0 & t_{\downarrow} & \zeta^{(3)}_{\uparrow\downarrow} & \zeta^{(3)}_{\downarrow\downarrow}  &\\
. & . & . & . & . & . & . \\
\end{array}
 \right),
\end{equation}
which is Hermitian and of size $2N\times2N$.
Here, the diagonal matrix elements~$\zeta^{(i)}_{\sigma\sigma}$ have the form
\begin{eqnarray}
\zeta^{(i)}_{\sigma\sigma}(i\omega_n) = i\omega_n + \mu_\sigma - V_i - \Sigma_{\sigma\sigma}^{(i)}(i\omega_n),
\end{eqnarray}
and the off-diagonal elements $\zeta^{(i)}_{\uparrow\downarrow}(i\omega_n)$ are defined accordingly to Eq.~(\ref{eq.15}).

As in the two-sublattice case, we close the DMFT-loop with the lattice Dyson equation,
\begin{equation}
  [{\cal G}^{(i)}(i\omega_n)]^{-1}_{\sigma_1\sigma_2} 
= [{\bf G}^{(i)}(i\omega_n)]^{-1}_{\sigma_1\sigma_2} + \Sigma_{\sigma_1\sigma_2}^{(i)}(i\omega_n),
\end{equation}
where ${\bf G}^{(i)}$ is a $2\times2$ block of the real-space Green's function matrix ${\bf G}$ obtained by inversion of (\ref{eq.17}). Finally, one defines a new set of Anderson parameters (as in the two-sublattice case) for each lattice site.

In case of a large system size (when the total number of lattice sites $N\gtrsim10^3$) the inversion of the real-space matrix~(\ref{eq.17}) becomes a time-consuming task in the numerical calculations (in comparison with the impurity solver with a moderate number of bath orbitals). 
Nevertheless, even with this limited total number of lattice sites R-DMFT is capable of a proper description of proximity-induced effects in lattice systems with magnetic order.

\section{Results}\label{sec.4}
\subsection{Homogeneous systems}
\subsubsection{Unpolarized mixtures with mass imbalance}\label{sec.4.1.1}
First, let us discuss the effects originating only from the mass imbalance in the system. We set $\mu_\uparrow=\mu_\downarrow$, which results in balanced populations ($P=0$) in a homogeneous system. According to Sec.~\ref{sec.2}, we note that the magnetic ground state of this system at half filling is a Z antiferromagnet (Z-AF) for any nonzero value of the mass imbalance.

\begin{figure}
\includegraphics{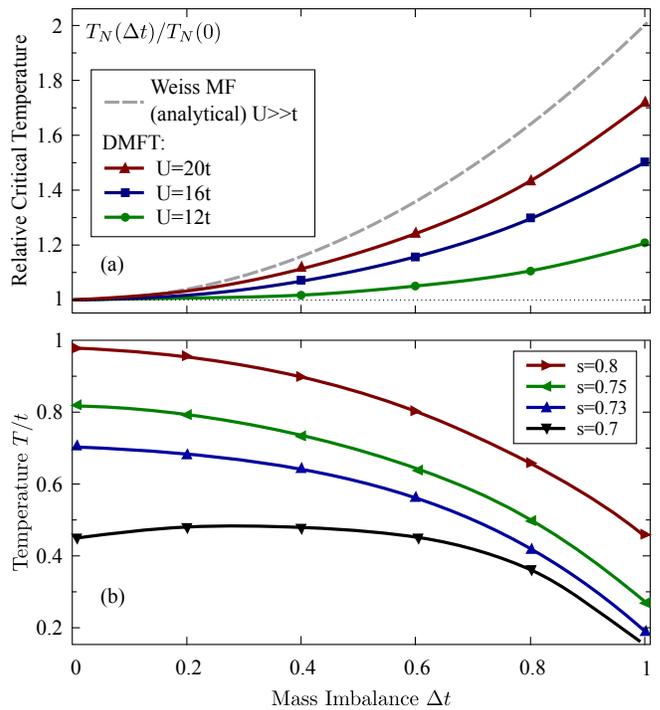}
\caption{\label{01crit} (Color online)
    (a) Critical temperature for N\'{e}el ordering in the half-filled Hubbard model ($\mu_{\uparrow,\downarrow}=U/2$) versus mass imbalance at different values of the interaction strength in a cubic lattice, obtained within DMFT. (b) Dependence of the minimal temperature that can be reached for a fixed entropy~$s$ per particle by minimizing over the interaction strength $U/t$, as a function of the mass-imbalance parameter $\Delta t$.
}
\end{figure}

A first important effect that we want to emphasize is that, according to mean-field analysis and existing Monte Carlo calculations in the limiting cases $\Delta t=0$ and $\Delta t=1$ (see Refs.~\cite{Sandvik1998PRL,Talapov1996JPA}, respectively), the critical (N\'{e}el) temperature increases with the mass imbalance. Indeed, by taking the mean-field definition for the N\'{e}el temperature in the Heisenberg model \cite{Ashcroft1976}, $T_{N} ={6JS(S+1)}/{3}$, where $S$ is the fermions' spin, one obtains for constant ${U}/{t}$, where $t=(t_\uparrow+t_\downarrow)/2$, and $J=J_{\parallel}$ [see Eq.~(\ref{eq.4})],
\begin{equation}
 {T_{N}(\Delta t)} = (1+\Delta t^2){T_{N}(0)}.
\end{equation}
In order to prove that this effect also takes place for the Hubbard model~(\ref{eq.1}) at moderately strong interactions, $U\gtrsim z t$, we numerically calculated $T_c$ by means of two-sublattice DMFT for a cubic lattice. The results are shown in Fig.~\ref{01crit}(a), which clearly confirms this behavior.
According to Ref.~\cite{Sotnikov2012PRL}, this phenomenon can be explained as caused by the suppression of quantum fluctuations (due to the emergence of an energy gap for one of the Goldstone modes) in systems with nonzero mass imbalance.

Another interesting effect concerns the entropy analysis for homogeneous systems. As pointed out in Ref.~\cite{Sotnikov2012PRL}, one can approach much closer the ordered state for mass-imbalanced mixtures with the same entropy values, compared to the balanced system. 
In order to show this, we present in Fig.~\ref{01crit}(b) the dependence of the minimal temperature that can be reached for constant entropy when minimizing over the interaction strength $U/J$ (thus assuming an  adiabatic change of the interaction strength) as a function of the mass-imbalance parameter.
Note that for $s=0.7$ this curve has a maximum. This is because for small mass-imbalance the lowest temperature is reached for large $U/t$ such that the system is in a Mott insulator. In this regime the minimal temperature increases when the mass imbalance is made larger. For even larger mass imbalance the lowest temperature is reached at small $U/t$ such that the system is in the Fermi liquid phase. In this phase the lowest temperature decreases with the mass imbalance, giving rise to the observed maximum.

It is important to note that it is now well established that DMFT quantitatively overestimates the critical temperature and critical entropy values in comparison with more exact methods for balanced mixtures. According to our analysis, this is also the case for systems with nonzero mass imbalance.
However, we should stress that the DMFT approach becomes more precise in the limit of large mass imbalance not only regarding the quantitative estimates of the critical temperature, but also in the critical entropy, the value of which does not depend on mass imbalance within the dynamical mean-field description for $U>zt$, $s^{\rm mf}_c=\ln2\approx0.69$. This can be seen, in particular, from the comparison with known results for the Heisenberg model \cite{Wessel2010PRB} ($U\gg t$, $\Delta t \rightarrow0$), where $s_c\approx0.34$, and results for the Ising model \cite{Sykes1972JPA} ($U\gg t$, $\Delta t\rightarrow1$), where $s_c\approx0.56$. Hence, the discussed advantages of imbalanced mixtures should be even more pronounced in studies based on more exact methods.

It should be noted that in addition to the spin-density wave in mass-imbalanced mixtures one also observes a weak charge-density wave (CDW) modulation. This is directly related to the presence of both N\'{e}el ordering and mass imbalance in the system and corresponds to the fact that in the Z-AF state the sites occupied by a heavier component have an enhanced double occupancy due to hopping from adjacent sites, which are occupied by the lighter component (in the same way, the opposite mechanism works for the sites occupied by a lighter component). The magnitude of this CDW, according to the estimates presented in Ref.~\cite{He2012PRA} for $U>zt$, is proportional to $\Delta t(U/t)^{-2}$. Hence, the CDW is more pronounced at moderate interaction strength, and it vanishes in the large $U/t$ limit.

\subsubsection{Polarized mixtures with mass imbalance}
\begin{figure}
\includegraphics{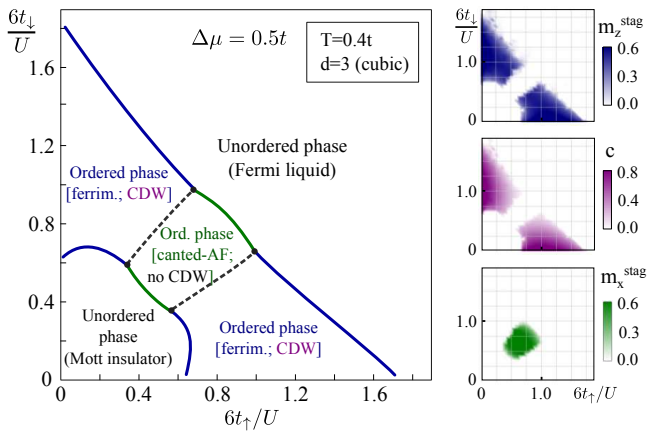}
    \caption{\label{02tatb} (Color online)
    Phase diagram for a population- and mass-imbalanced mixture at half filling and finite temperature, obtained within DMFT for a cubic lattice. Dashed and solid lines correspond to the first- and second-order phase transition lines, respectively. The CDW parameter is defined in terms of the double occupancy $D_i=\langle\hat{n}_{i\uparrow}\hat{n}_{i\downarrow}\rangle$, $c=(D_i-D_{i+1})/(D_i+D_{i+1})$; $m_{z,x}^{\rm stag}=\langle\hat{S}_i^{Z,X}\rangle-\langle\hat{S}_{i+1}^{Z,X}\rangle$.
    }
\end{figure}

The presence of both population and mass imbalance in ultracold mixtures according to the effective Hamiltonian~(\ref{eq.3}) results in competition between different types of antiferromagnetic ordering. In Fig.~\ref{02tatb} we present the phase diagram showing the structure of the ordered phases in the intermediate parameter region at half filling. [For moderate polarizations $P<0.8$ and $U>zt$, the condition of half filling is fulfilled by taking the average chemical potential $\bar{\mu}=U/2$, where $\bar{\mu}\equiv(\mu_\uparrow+\mu_\downarrow)/2$, but for larger population imbalances it must be  adjusted by hand.]
When the hopping amplitude of both species is low in comparison with the interaction strength, the system is in a paramagnetic Mott insulating state. In the opposite case, the system is in the unordered Fermi liquid phase, which is compressible and characterized by an enhanced double occupancy of the lattice sites. In the central region, two different phases appear: {\it a canted antiferromagnet}, which is characterized by a staggered magnetization in the $X$ direction and a net magnetization in the $Z$ direction, and {\it a ferrimagnet} (labeled by Z-AF in this diagram), which has both a staggered and a net magnetization in the $Z$ direction. 
Note that in the canted-AF phase, the occupation of each site by heavy and light species is equal; thus, an additional CDW does not appear in this case. In the region of intermediate mass imbalance, a first-order phase transition (with a narrow coexistence region) takes place between the phases with different magnetic order.

\begin{figure}
\includegraphics{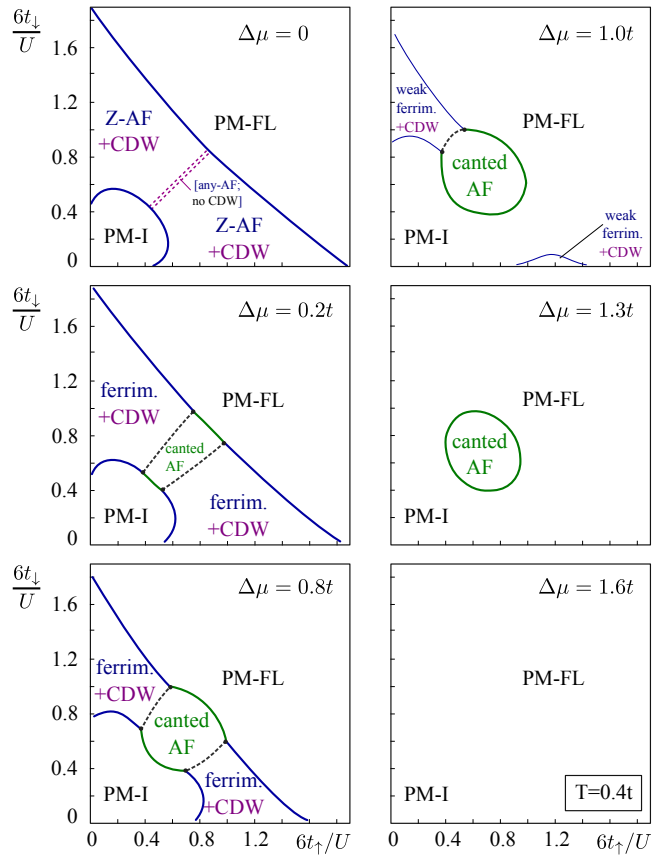}
    \caption{\label{03evol} (Color online)
    Set of phase diagrams for a population- and mass-imbalanced mixture at half filling, $T=0.4t$, and different $\Delta\mu$, obtained within DMFT for a cubic lattice, including easy-axis antiferromagnetic (Z-AF), charge-density wave (CDW), ferrimagnetic, canted AF, unordered Mott insulator (PM-I), and unordered Fermi liquid (PM-FL) phases.
    }
\end{figure}

We now discuss how the phase diagram presented in Fig.~\ref{02tatb} changes when $\Delta\mu$ (which defines the polarization~$P$) is increased
(see also Fig.~\ref{03evol}).
At $\Delta\mu =0$ the whole region with magnetic order corresponds to the Z-AF phase, except the central part (namely, the diagonal line $t_\uparrow = t_\downarrow$), where the AF order can be set in any direction. When population imbalance appears, the central phase characterized by  canted order develops. This phase increases in width (towards the regions with large mass imbalance) when $\Delta\mu$ becomes larger. As for the ferrimagnetic region, it shrinks not only from the side of the canted-AF phase, but also from the sides of unordered phases, as the net magnetization in $Z$ direction suppresses weak Z-AF order in the region close to phase boundaries.
Close to the critical value of the chemical potentials difference ($\Delta\mu=1.0t$ in Fig.~\ref{03evol}) we observe an asymmetry in the vanishing of weak ferrimagnetic phases in the phase diagram. We see that the ferrimagnetic phase is slightly more stable, when one one realizes a system with a majority of heavy particles. One can understand this in a way that weak magnetic correlations are less suppressed by excitations in this region due to a lower kinetic energy of the remaining uncorrelated fraction of the system.
At large population imbalances ($\Delta\mu\gtrsim1.3t$ for $T=0.4t$) the ferrimagnetic phase completely vanishes, leaving only the region with canted-AF order in the central part. 

\begin{figure}
\includegraphics{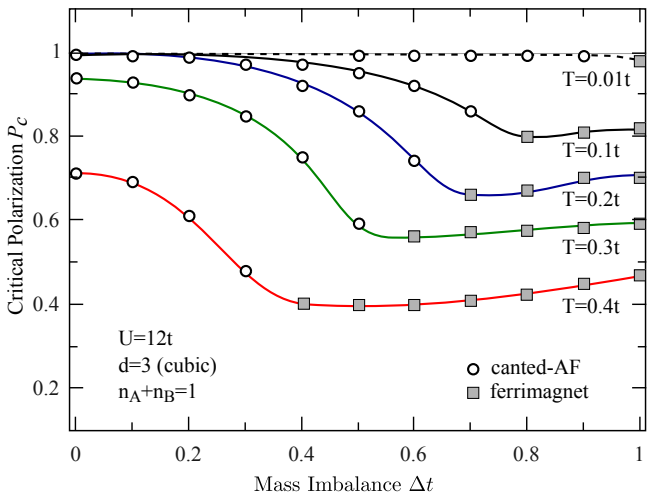}
    \caption{\label{03trip} (Color online)
    Dependence of critical polarization destroying AF order on the mass-imbalance parameter, obtained by DMFT.
    }
\end{figure}

The antiferromagnetic order is therefore at a given temperature most unstable against imbalances in the triple-point region of a diagram of the type presented in Figs.~\ref{02tatb} and \ref{03evol}, where canted, ferrimagnetic, and paramagnetic phases coexist. In Fig.~\ref{03trip} we show this effect for different temperatures of the imbalanced mixtures.
The critical polarization clearly reveals the sensitivity of the magnetically ordered phases to an increase of imbalance in the triple-point regions, where corresponding minima are observed. Also, the effect that ferrimagnetic states are suppressed by population imbalance can be seen from this plot by comparing the left-hand (canted) and right-hand (ferrimagnetic) sides of curves presented in Fig.~\ref{03trip}.
Interestingly, DMFT predicts that at zero temperature antiferromagnetic order develops for any polarization. This is different for the case of attractive interactions and superfluidity, which is destroyed at a finite Chandrasekhar-Clogston limit for the polarization \cite{Chandrasekhar62, Clogston62, Lobo06}, which was beautifully demonstrated in experiments with ultracold Fermi gases \cite{Shin08}.

\subsection{Finite systems in a harmonic trap}
Finally, we turn our analysis to the case where an additional external trapping potential is present in the system. For simplicity, below we assume that the trapping potential~$V_i$ in Eq.~(\ref{eq.1}) has an axial-symmetric form, $V_i = V_0 r_i^2/a^2$, where $V_0$ is the strength of the harmonic potential, $r_i$ is the distance from the lattice site~$i$ to the trap center, and $a$ is the lattice constant. Evidently, all the results of this section can be extended to include anisotropies usually present in real experiments.

We start our analysis by applying the LDA in combination with two-sublattice DMFT introduced in Sec.~\ref{sec.3.2}. Within LDA, we perform calculations for a particular lattice site by taking $\mu^{(i)}_{\sigma} = \mu^{(0)}_{\sigma} + V_i$ in the DMFT calculations. In all results presented in this section the average chemical potential in the center of the trap is taken as $(\mu^{(0)}_\uparrow + \mu^{(0)}_\downarrow)/2 = U/2$, such that the system is at half filling in the trap center [at least  for $|\Delta\mu|\ll U$ ($\Delta\mu = \mu^{(0)}_\uparrow - \mu^{(0)}_\downarrow$)], which is the most relevant case for our investigation. 

\begin{figure}
\includegraphics{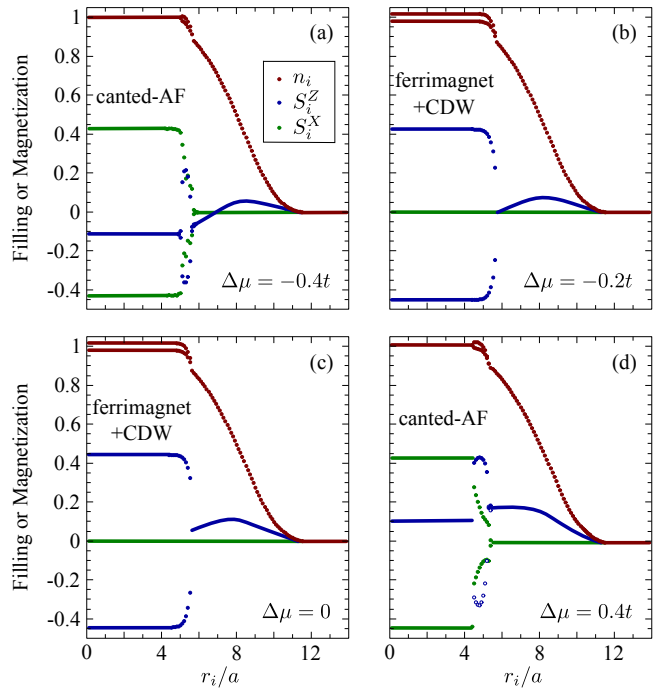}
    \caption{\label{04rlda} (Color online)
    Real-space density and magnetization distributions for a cubic lattice in a harmonic trap obtained by LDA+DMFT. Parameters used in (a)--(d): $U=12t$, $T=0.2t$, $\Delta t = 0.2$, and $V_i=0.1t(r_i/a)^2$, where $a$ is the lattice constant.}
\end{figure}

Performing calculations for  fixed mass imbalance and varying chemical potential differences $\Delta\mu$ in the trap center, we obtain distributions of the total filling and magnetization in various directions, which are presented in Fig.~\ref{04rlda}.
There are several details in those plots that are worth discussing. First of all, we note that for nonzero mass imbalance and $\Delta\mu=0$ one has a (globally) polarized mixture due to a ``ferromagnetic'' shell originating from a wider distribution of the lighter component in a trap. Hence, for the purpose of obtaining Z-AF order in the trap center, one should account for this effect and adjust the total polarization (or make the harmonic trap also species dependent). 
One could also use this adjustment in a combination with the additional cooling mechanism in mass-imbalanced mixtures, pointed out in Ref.~\cite{Winograd2012PRB} (removing from the system a part of the heavy component that carries a larger amount of entropy at $U=0$).
Note, however, that as long as one only specifies the chemical potential difference, there is no problem here.

Possible advantages of the Z-AF phase could be not only the issues related to cooling as pointed out in Ref.~\cite{Sotnikov2012PRL} and in Sec.~\ref{sec.4.1.1}, but also a more convenient detection, e.g., by single-site resolution imaging with a quantum gas microscope \cite{Bakr2009Nat}, where the alternating density of one spin (species) component can be directly detected. In contrast, in the canted-AF phase, the density of the spin components does not alternate from site to site, thus one has to analyze the behavior of the double occupancy \cite{Gor2010PRL,Fuchs2011PRL}, merge nearest-neighbor lattice sites \cite{Greif2012pre} or introduce additional methods to detect nearest-neighbor spin-spin correlation functions in the $XY$ plane, such as Bragg spectroscopy \cite{Corcovilos2010PRA} or noise correlations analysis \cite{Bruun2009PRA}.

It is worth noting that the CDW structure peculiar to the Z-AF phase is clearly seen in the distributions of the total particle number in Figs.~\ref{04rlda}(b) and \ref{04rlda}(c) (double lines in the central part), while it is absent in Figs.~\ref{04rlda}(a) and \ref{04rlda}(d), where the canted-AF phase developed in the bulk. As one sees, the shell structure also has an interesting dependence on the population imbalance. The magnetization in it increases with total polarization (adding light and removing heavy particles from the system), but with the decrease of total polarization the ``ferromagnetic'' shell related to mass imbalance does not vanish. Instead, according to Fig.~\ref{04rlda}(a), a double-shell structure develops with inner and outer ``ferromagnetic'' shells originating from population and mass imbalances, respectively.

\begin{figure}
\includegraphics{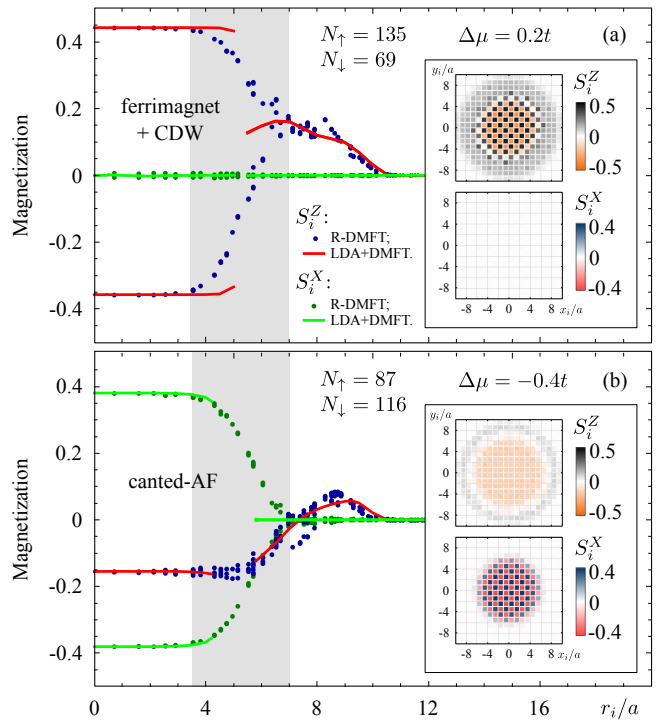}
    \caption{\label{05rdft} (Color online)
    Comparison of real-space magnetization profiles for a square lattice in a harmonic trap obtained by LDA+DMFT (lines) and R-DMFT (dots). Shadowed areas correspond to the regions where LDA fails to reproduce the detailed structure and has problems with convergence. 
    (Insets) Contour plots representing real-space distributions of magnetizations obtained by R-DMFT with the same parameters.
    Parameters used in (a) and (b) and insets: $U=12t$, $T=0.2t$, $\Delta t = 0.2$, and $V_i=0.1t(r_i/a)^2$.
    }
\end{figure}

There are also discontinuities present in the magnetization close to the boundaries of the antiferromagnetic phases: Since the applied LDA+DMFT approach does not include proximity effects these sharp features are not smeared out by the trap. Moreover, in these regions this approach has a rather bad convergence.
In Fig.~\ref{05rdft}
we show that by accounting for the proximity effect within R-DMFT, the magnetization in trapped imbalanced mixtures has a smooth behavior with a larger region of stability of the antiferromagnetic phase than predicted by LDA.

\section{Conclusions}
We studied antiferromagnetically ordered phases that emerge in two-component ultracold fermionic mixtures with mass  and population imbalance. 
Our analysis was based on DMFT and its real-space generalization at finite temperature.

It is pointed out that two types of imbalance favor different types of antiferromagnetically ordered phases: The ferrimagnetic phase is favored by mass imbalance, while the canted-antiferromagnetic phase is favored by population imbalance.
In the absence of population imbalance, we demonstrated within DMFT the advantages of mass-imbalanced mixtures, i.e., an increase of the critical temperature and the possibility to cool the system by adiabatic change of the interaction strength to lower temperatures than possible for balanced mixtures.
Guided by the exact values of the critical entropy in the limiting cases of the Heisenberg and Ising models, we argued that DMFT gives a better prediction for the critical entropy for imbalanced mixtures compared to balanced ones, which makes systems with mass imbalance even more advantageous for the purpose of observing magnetic ordering phenomena in optical lattices.

In the presence of both mass and population imbalance we obtained the finite-temperature phase diagram with the corresponding first-order phase transition between the different AF states. We revealed that AF order is most unstable against thermal fluctuations in the triple-point regions. To this end, we performed a stability analysis of the ordered phases against population imbalance at different temperatures. At zero temperature we found that for all polarizations antiferromagnetic order develops.

We also obtained real-space density and magnetization distributions of imbalanced mixtures in a harmonic trap. It is shown that, depending on the total polarization, the mass-imbalanced mixture can have different ordered phases in the bulk and different magnetic shell structures. The detailed description of these effects could help not only in preparing the mixture closely to its equilibrium state, but also in the detection of antiferromagnetic correlations in ultracold gases.

\begin{acknowledgments}
The authors thank to D. Cocks, A. Georges and W. Zwerger for useful discussions.
Support by the German Science Foundation DFG via Sonderforschungsbereich SFB/TR 49 and Forschergruppe FOR 801 is gratefully acknowledged.
\end{acknowledgments}

\bibliography{A15}

\end{document}